\documentclass[pra,twocolumn]{revtex4}
\usepackage{amsmath,amssymb,graphics}
\newcommand{\bra}[1]{\langle #1 |}
\newcommand{\ket}[1]{| #1 \rangle}
\newcommand{\dyad}[2]{\ket{ #1 } \! \bra{ #2 }}

\newcommand{\hx}{\hat{x}}
\newcommand{\hp}{\hat{p}}

\newcommand{\vp}{\mathbf{p}}
\newcommand{\vx}{\mathbf{x}}
\newcommand{\vq}{\mathbf{q}}
\newcommand{\vc}{\mathbf{c}}
\newcommand{\xcl}{\overline{x}}
\newcommand{\qcl}{\overline{q}}
\newcommand{\cphi}{\tilde\phi}
\newcommand{\cqcl}{\tilde{q}}
\newcommand{\cS}{\tilde S}
\newcommand{\vxcl}{\mathbf{\overline{x}}}
\newcommand{\vpcl}{\mathbf{\overline{p}}}
\newcommand{\vqcl}{\mathbf{\overline{q}}}
\newcommand{\vcqcl}{\mathbf{\tilde{q}}}
\newcommand{\vcxcl}{\mathbf{\tilde{x}}}
\newcommand{\vcpcl}{\mathbf{\tilde{p}}}
\newcommand{\vF}{\mathbf{F}}
\newcommand{\hvp}{\hat{\vp}}
\newcommand{\hvx}{\hat{\vx}}
\newcommand{\hvq}{\hat{\vq}}

\newcommand{\hU}{\hat{U}}

\newcommand{\hD}{\hat{D}}
\newcommand{\hW}{\hat{W}}
\newcommand{\hH}{\hat{H}}
\newcommand{\kH}{\mathcal{H}}
\newcommand{\Jobj}{\mathcal{J}}
\newcommand{\Tr}{\text{Tr}}
\newcommand{\Id}{\mathbb{I}}
\newcommand{\kB}{k_\mathrm{B}}
\newcommand{\wilcox}{\eta}
\newcommand{\aosc}{a_0}
\begin{document}
\title{Correcting errors in a quantum gate with pushed ions via
  optimal control}
\author{Uffe V. Poulsen}
\affiliation{Lundbeck Foundation Theoretical
  Center for Quantum System Research\\
  Department of Physics and Astronomy,
  University of
  Aarhus,
  DK 8000 Aarhus C, Denmark}
\author{Shlomo Sklarz}
\author{David Tannor}
\affiliation{Department of Chemical Physics,
Weizmann Institute of Science,
76100 Rehovot , Israel}
\author{Tommaso Calarco}
\affiliation{Institut f\"{u}r
Quanteninformationsverarbeitung,
Albert-Einstein-Allee 11,
D-89069 Ulm,
Deutschland\\
European Centre for Theoretical Studies in Nuclear Physics and Related Areas, I-38050 Villazzano (TN), Italy}
\begin{abstract}
  We analyze in detail the so-called ``pushing gate'' for trapped
  ions, introducing a time dependent harmonic approximation for the
  external motion. We show how to extract the average fidelity for the
  gate from the resulting semi-classical simulations. We characterize
  and quantify precisely all types of errors coming from the quantum
  dynamics and reveal for the first time that slight nonlinearities in
  the ion-pushing force can have a dramatic effect on the adiabaticity
  of gate operation. By means of quantum optimal control techniques,
  we show how to suppress each of the resulting gate errors in order
  to reach a high fidelity compatible with scalable fault-tolerant
  quantum computing.
\end{abstract}
\maketitle
\section{Introduction}
\label{sec:intro}
Trapped ultracold ions have represented a major candidate for the
implementation of scalable quantum information processing since the
beginning of this research field. The first proposal of an ion-based
quantum computer by Cirac and Zoller in 1995 \cite{CZ95} has been
followed by a great variety of other schemes based on ions
\cite{CZ00}, on other
quantum optical systems like neutral atoms \cite{Jaksch99,Chip00} and
on solid-state systems \cite{LossDiVincenzo99} as well. With the
progress of experimental techniques and the demonstration of
entangling quantum gates based on several different candidate physical
systems, the focus has progressively shifted toward the fulfilment of
scalability desiderata \cite{divincenzo00:_physic_implem_qc}, that is,
the realization of quantum gates with very high fidelities, in the range
0.999 -- 0.9999.

Gate errors in a real implementation of a given quantum gate scheme
can be reduced by different means. Some errors arise from (or are
increased by) experimental imprecisions of a technical nature and can
be controlled by careful alignment, stabilization etc.\ of the
experimental apparatus. Other errors stem from unaviodable
interactions with the environment and can be reduced by simply
completing the gate in as short a time as possible. Typically, a gate
scheme can be made faster by simple scaling to higher intensities,
shorter distances etc.\ If such simple optimizations of the gate prove
insufficient, one needs to consider changes to the scheme itself and
trade simplicity for improved performance. This is exactly the goal of
\emph{quantum optimal control techniques}
\cite{dalessandro08:_intro_quant_contr_dyn} which allow for a precise
tailoring of the system's evolution by time-dependent tuning of some
external parameters. With sufficient control over these parameters, a
given target state can often be reached with minimal errors even over
short gate operation times. The application of these methods to
quantum information systems requires in turn a very accurate
simulation of the dynamics and a careful understanding of the targeted
error sources. This is precisely the aim of this paper, in the
specific case of the two-qubit ion gate first proposed in \cite{CZ00}
and subsequently analyzed in
\cite{calarco01:_entan_micro_traps,sasura03:_fast_ql_selec_displace}. In
this ``pushing gate'' the qubits are encoded in the internal states of
two ions. Each ion is held in a separate microtrap and state-selective
``push'' potentials are applied in order to modify the distance and
thus the Coulomb interaction between ions, see
Sec.~\ref{sec:pushing_gate} below.  We shall first point out a series
of issues that arise when the assumption of spatial homogeneity of the
ion-driving force is dropped, and subsequently develop a way to
correct each of these issues, exploting a range of ideas, including in
a crucial way optimal control methods.

It should be noted from the outset that in the present paper, we
analyze the pushing gate without what is called the $\pi$-pulse method
in
Refs.~\cite{calarco01:_entan_micro_traps,sasura03:_fast_ql_selec_displace},
where it was shown to dramatically reduce some types of errors. The
$\pi$-pulse method is a spin-echo technique and requires the gate to
be repeated with the internal state of the ions flipped. Typically,
single particle operations like flipping the internal states can be
done with high fidelity, and it is reasonable to expect that
eventually the $\pi$-pulse method will be used. However, internal
state control is at least in principle a separate issue from the
pushing gate operation itself. Keeping the design process modular in
spirit, it is relevant to optimize the gate without this additional
trick and thus pave an alternative road to high fidelities. As will
become clear below, we extract quite general noise reduction methods
from the automated numerical optimization results and it is an
interesting topic for future research to combine these with the
$\pi$-pulse method.

The remainder of the paper is organized as follows. In
Sec.~\ref{sec:td-potentials} we introduce the general setting of
conditional dynamics gate schemes, a useful approximation for
simulating such a gate, and a measure of the gate errors. In
Sec.~\ref{sec:pushing_gate} we specialize to the pushing gate. The
un-optimized performance of the gate is reported in
Sec.~\ref{sec:results} and in Sec.~\ref{sec:optimal_control} we show
how this performance can be significantly improved by a combination of
``manual'' changes and numerical optimal control methods. Finally we
conclude in Sec.~\ref{sec:conclusion}. The Appendix contains a number
of more technical results and derivations. 

\section{Conditional dynamics}
\label{sec:td-potentials}
The basic idea of the so-called pushing gate is one of \emph{conditional
  dynamics}, i.e.\ we apply potentials depending on the internal state
of the two ions. The internal states themselves are not changed during
the gate, or, in the case of the $\pi$-pulse method,
are changed on a much shorter time-scale than the external dynamics.
This means that the analysis of the problem splits into four separate
evolutions for the external state, one for each of the logical
(internal) states, $00$, $01$, $10$, and $11$. In the following, these
four evolutions will be denoted ``branches'' and will be indexed by
$\beta\in\{00,01,10,11\}$. The complete Hamiltonian can be written in
the form of a sum of internal$\otimes$external factorized terms
\begin{equation}
  \label{eq:cond_dyn_hamiltonian}
  \hH^\text{tot}(t)
  =
  \sum_\beta \dyad{\beta}{\beta}\otimes \hH^\text{ex}_\beta(t)
\end{equation}
and it results in an evolution operator of a similar form
\begin{equation}
  \label{eq:con_dyn_U}
  \hU^\text{tot}(t)
  =
  \sum_{\beta} \dyad{\beta}{\beta}
  \otimes
  \hU_{\beta}^\text{ex}(t)
  .
\end{equation}

Ideally, when $t=T$ at the end of the gate, the $U_\beta^\text{ex}$ should differ from each other by at
most a phase factor multiplying a common unitary operator $\hU^\text{ex}_\text{com}$
\begin{equation}
  \label{eq:u_ex_phase_factor}
  \hU^\text{ex}_\beta(T)
  =
  e^{i\theta_\beta}\hU^\text{ex}_\text{com}
\end{equation}
so that $\hU^\text{tot}$ itself can be factorized:
\begin{equation}
  \label{eq:utot_factor}
  \hU^\text{tot}(T)
  =
  \left(
    \sum_\beta \dyad{\beta}{\beta}e^{i\phi_\beta}
  \right)
  \otimes
  \hU^\text{ex}_\text{com}
  .
\end{equation}
The internal evolution is then that of a phase gate, while the
external evolution can in principle be undone using internal-state
independent potentials.

The requirement (\ref{eq:u_ex_phase_factor}) is very hard to achieve
and would make the gate completely independent on the initial external
state. However, we can typically assume to have some degree of control
over the initial external state, e.g.\ by cooling the particles before
the gate. This means that~(\ref{eq:u_ex_phase_factor}) need only hold
when restricted to a subset of the complete Hilbert space, typically
the states of relatively low energy. In Appendix~\ref{sec:fidelity} we
show how to evaluate the performance of the gate in general. For now,
we note that since only the low energy part of
$\hU^\text{ex}_\beta(T)$ will be important, we can focus on getting a
good approximation for this part when trying to simulate the gate
dynamics.

A low initial energy means particles localized near the potential
minimum and this suggests using a harmonic approximation to the real
potential~\endnote{For a thorough introduction to semi-classical wave
  packet methods, see e.g.~
  Ref.~\cite{littlejohn86:_semiclas_evo_wave_packets}.}. The simplest
choice is to Taylor expand around a fixed point, which is
not changed during the gate operation. The next level of refinement is
to expand around the instantaneous potential minimum. This works very
well if the gate operation is nearly adiabatic so the particles stay
near the (moving) minimum at all times.  However, it may be desirable
to make fast and substantial changes to the potential during the gate
and that may induce pronounced non-adiabatic dynamics. In that case,
the harmonic approximation can still be a good one provided it is done
around the \emph{classical trajectories} of the particles. Typically
these trajectories cannot be computed analytically, but for any
moderate number of particles it is a numerically simple task to find
them. Below we will use this method and show how it leads to a
relatively simple characterization of $U^\text{ex}_\beta$.

\subsection{Harmonic approximation}
\label{sec:harmonic_approx}

In this section we focus on a single branch of the evolution and thus suppress the
$\beta$ index. Let us denote by $\vxcl(t)$
the classical trajectory, which is found by solving classical equations
of motion. The time-dependent, second order Taylor
expansion of the potential $V(t,\vx)$ around $\vxcl(t)$ reads
simply
\begin{equation}
  \label{eq:second_order_taylor}
  V_\text{so}(t,\vx)
  =
  V^{(0)}(t)
  +
  \Delta\vx^\mathrm{T}
  \mathbf{V}^{(1)}(t)
  +
  \frac{1}{2}
  \Delta\vx^\mathrm{T}
  V^{(2)}(t)
  \Delta\vx
  ,
\end{equation}
with $\Delta\vx=\left[\vx-\vxcl(t)\right]$ and
\begin{equation}
  \label{eq:def_taylor}
  \begin{split}
    V^{(0)}(t)
    &=
    V(t,\vxcl)
    ,
    \\
    V^{(1)}_i(t)
    &=
    \frac{\partial V}{\partial x_i} (t,\vxcl)
    ,
    \\
    V^{(2)}_{ij}(t)
    &=
    \frac{\partial^2 V}{\partial x_i \partial x_j} (t,\vxcl)
    .
  \end{split}
\end{equation}
Note that we will still use $\vx$ as our coordinate, i.e.\ we are \emph{not}
changing to a coordinate system moving with $\vxcl(t)$, we
simply use a potential that approximates the real potential close
to $\vxcl(t)$. Collecting terms of equal order in $\vx$ leads
to the alternative form
\begin{equation}
  \label{eq:so_taylor_order}
  V_\text{so}(t,\vx)
  =
  E(t)
  -
  \vx^\mathrm{T}
  \mathbf{F}(t)
  +
  \frac{1}{2}
  \vx^\mathrm{T}
  K(t)
  \vx
  ,
\end{equation}
with
\begin{equation}
  \label{eq:def_eff}
  \begin{split}
    E(t)
    &=
    V^{(0)}(t)-\vxcl^\mathrm{T}(t)\mathbf{V}^{(1)}(t)
    +\frac{1}{2}\vxcl^\mathrm{T}(t)V^{(2)}(t)\vxcl(t)
    ,
    \\
    \mathbf{F}(t)
    &=
    -\mathbf{V}^{(1)}(t)+{V}^{(2)}(t)\vxcl(t)
    ,
    \\
    {K}(t)
    &=
    {V}^{(2)}(t)
    .
  \end{split}
\end{equation}

\subsection{Gaussian evolution}
\label{sec:gauss_evol}

The big advantage of choosing a second order approximation to the real
potential is that this restricts the corresponding approximate
$\hU^\text{ex}(t)$ to be \emph{Gaussian} for all $t$. Let us introduce
the compact notation $\vq=(\vx,\vp)$ and define the matrix $J$ by 
\begin{equation}
  \label{eq:def_J}
  J
  =
  \begin{bmatrix}
    0 & \Id_n \\
    -\Id_n &0
  \end{bmatrix}
  ,
\end{equation}
where $n$ is the number of degrees of freedom.
Then the usual canonical commutation relations can be written
as
\begin{equation}
  \label{eq:commu_relations}
  [q_i,q_j] = i J_{ij}
  .
\end{equation}
With the potential of Eq.~(\ref{eq:so_taylor_order}) and a matrix of
particle masses $M=\text{diag}(m_1,\ldots,m_n)$ the time-dependent
Hamiltonian becomes
\begin{equation}
  \label{eq:so_Ham}
  \hH_\text{so}
  =
  \frac{1}{2} \hvp^\mathrm{T}M^{-1}\hvp+V_\text{so}(t,\hvx)
  .
\end{equation}
In Appendix~\ref{sec:evol_quad_ham} we show that such a Hamiltonian
leads to an evolution operator of the form
\begin{equation}
  \label{eq:evol_operator}
  \hU(t)
  =
  e^{ -i \phi(t)}
  \hD_{\vc(t)}
  \hW_{b(t)}
  ,
\end{equation}
where $\hD_\vc$ is a displacement operator and $\hW_b$ a squeezing
operator
\begin{equation}
  \label{eq:def_D_W}
  \begin{split}
  \hD_\vc &= e^{-i\mathbf{\vc}^\mathrm{T}J\hvq}
  \\
  \hW_b   &= e^{-i\frac{1}{2}\hvq^\mathrm{T} b \hvq}
  .
\end{split}
\end{equation}
The scalar $\phi$, the vector $\vc$, and the matrix $S=\exp(Jb)$
should satisfy the following equations of motion:
\begin{equation}
  \label{eq:eoms_d_b}
  \begin{split}
    \frac{\partial}{\partial t} \phi
    &= E -\frac{1}{2}\mathbf{F}^\mathrm{T}\vxcl
    ,
    \\
    \frac{\partial}{\partial t} \vc
    &=
    J h \vc
    +\begin{bmatrix}
      0\\
      \mathbf{F}
    \end{bmatrix}
    ,
    \\
    \frac{\partial}{\partial t} S &= J h S
    ,
  \end{split}
\end{equation}
where the $2n\times2n$ matrix $h$ is defined by
\begin{equation}
  \label{eq:def_h}
  h(t)=
  \begin{bmatrix}
    K(t) & 0 \\
    0 & M^{-1}
  \end{bmatrix}
  .
\end{equation}

The form of solution (\ref{eq:evol_operator})--(\ref{eq:eoms_d_b})
holds for any second order Hamiltonian. In the particular case where
$V_\text{so}$ is a Taylor expansion of a real potential around the
classical trajectory $\vxcl(t)$, the equation of motion for $\vc$
reduces to the exact equation of motion for $\vqcl=(\vxcl,\vpcl)$
where $\vpcl$ is the classical momentum. In the following, we will
therefore write $\vqcl$ instead of $\vc$. It is then important to
remember that the right-hand sides of Eqs.~(\ref{eq:eoms_d_b}) are in
general \emph{non-linear} functions of $\vxcl$.

\subsection{Fidelity}
\label{sec:what_to_plot}
We can quantify the performance of the gate by calculating the average
\emph{fidelity} $F_\mathrm{avg}$ (see Appendix~\ref{sec:fidelity})
between the obtained output state and the ideal one when the input
state is varied. One can then separate out three kinds of contributions to
the deviation of $F_\mathrm{avg}$ from 1 (the perfect gate)
\begin{equation}
  \label{eq:fidelity_split}
  1-F_\mathrm{avg}
  =
  \mathrm{E}_\theta
  +
  \mathrm{E}_\vqcl
  +
  \mathrm{E}_S
\end{equation}
The three types of errors each have their physical interpretation. The
most straightforward one pertains to the \emph{sloshing errors}
$\mathrm{E}_\vqcl$ which correspond to a residual motion of the ions
after the gate has been completed and the micro-traps are again at
rest. The \emph{phase errors} $\mathrm{E}_\theta$ are errors in the gate
phase. Finally, the \emph{breathing errors} $\mathrm{E}_S$ are induced by
differences in the harmonic approximation parameters around the
classical trajectory for different internal states. For example, in
the case we consider, when the particles are pushed closer together,
the second order term in the Coulomb repulsion becomes larger,
cf. Eq.~(\ref{eq:K_ii}) below.

In our model we assume
that systematic, \emph{local} phase errors can be undone. 
Then an explicit calculation in Appendix~\ref{sec:fidelity}, shows that 
the phase errors are given by
\begin{equation}
  \label{eq:def_phase_errors}
  \mathrm{E}_\theta
    =
    \frac{1}{20}
    \left(
      \theta_{00} - \theta_{01}
      -\theta_{10} +  \theta_{11}
      -\pi
    \right)^2
\end{equation}
with $\theta_\beta=-\phi_\beta+\Tr[b_\beta\gamma]$, where $\gamma$ is
the covariance matrix of the external state, see
Section~\ref{sec:Gaussian_case} of the Appendix.  Note the inclusion
of the $\Tr[b_\beta\gamma]$ terms in the definition of
$\theta_\beta$: These terms correspond to phase contributions from
average excitation energy in the traps and are therefore temperature
dependent through $\gamma$. For our parameters they are small. 

The sloshing errors are given by
\begin{equation}
  \label{eq:def_slosh_errors}
  \mathrm{E}_\vqcl
  =
  \frac{1}{20}\sum_{\alpha<\beta}
  (\vqcl_\alpha-\vqcl_\beta)^\mathrm{T}\gamma(\vqcl_\alpha-\vqcl_\beta)
  ,
\end{equation}
and the breathing errors are
\begin{equation}
  \label{eq:def_squeeze_errors}
  \begin{split}
  \mathrm{E}_S
  =&
  \frac{1}{40}\sum_{\alpha<\beta}
  \Tr \left[(b_\alpha - b_\beta) \gamma
    (b_\alpha - b_\beta) \gamma\right]
  \\
  &+\frac{1}{160}\sum_{\alpha<\beta}
  \Tr \left[(b_\alpha - b_\beta) J(b_\alpha - b_\beta) J
  \right]
  .
\end{split}
\end{equation}
Here $\phi_\beta$, $\vqcl_\beta$, and $b_\beta$ are defined in
Eqs.~(\ref{eq:evol_operator},\ref{eq:def_D_W}), and are the variables
describing the Gaussian approximation in the $\beta$ branch of the
evolution. Again $\gamma$ introduce temperature dependence. 

\section{The pushing gate}
\label{sec:pushing_gate}
Let us now focus on the particular case of the pushing gate. Here we
have two ions, each in a separate micro-trap~\endnote{The gate can
  also be implemented in a string of ions (see
  \cite{calarco01:_entan_micro_traps}) and treated by the method
  described in this paper, but we focus exclusively on the two ion
  case.}. To further simplify the discussion, we concentrate on just
one spatial dimension, i.e.\ there are two degrees of freedom,
$n=2$. The ions are assumed to be of identical mass, $m_1=m_2=m$, and
thus $M=m\Id_2$. The potential energy consists of a micro-trap for
each ion, time- and internal state-dependent pushing potentials, and
the Coulomb interaction. The pushing potentials can be realised as
optical dipole potentials generated by focused laser beams. The
time-dependence of these potentials is most easily achieved by
controlling the intensity of the laser and the state-selectivity by
polarization selection
rules~\cite{sasura03:_fast_ql_selec_displace}. We assume that the form
for the internal state labeled by $\beta$ is
\begin{equation}
  \label{eq:def_V}
  \begin{split}
  V^{(\beta)}(t,\vx)
  =&
  \sum_{i=1,2} \frac{1}{2}m\omega^2 x_i^2
  +
  \frac{e^2}{4\pi\epsilon_0 \left|d+x_2-x_1\right|}
  \\
  &+
  \sum_{i=1,2} f_i^{(\beta)}(t)\frac{\hbar\omega}{\aosc}
  \left[- (-1)^i   x_i + \frac{G}{\aosc} x_i^2
  \right]
  .
\end{split}
\end{equation}
The trapping potentials are assumed to be perfectly harmonic.  The
state-dependent pushing amplitudes $f^{(\beta)}_i$ are such that the
ions are only pushed if they are in the internal state 1,
$f^{(\beta)}_i(t) = \delta_{\beta_i,1}f(t)$. Note that the $(-1)^i$
factor means that ion 1 is pushed to the left and ion 2 to the
right. Non-linear contributions to the pushing potentials are included
\emph{via} the constant $G$. The harmonic oscillator ground state size
is $\aosc=\sqrt{\hbar/m\omega}$. The two coordinates $x_1$ and $x_2$
are taken to have origin in the respective trap centers, a distance
$d$ apart. Experimentally, the parameters in (\ref{eq:def_V}) can be
varied quite a lot (see
e.g.~\cite{sasura03:_fast_ql_selec_displace}). Trap distances $d$ from
$100\mu m$ all the way down to $1 \mu m$ are within technological
reach.  The trap frequency $\omega$ can be chosen in the range
$2\pi\times 10^4-10^7 \mathrm{Hz}$ which for e.g.\ Ca$^+$ ions will
mean an oscillator length $a$ from $150\mathrm{nm}$ down to
$5\mathrm{nm}$.

It should be noted that the use of optical dipole potentials to
generate the state selective pushing forces will in general introduce
large single qubit phases due to ac-Stark shifts. This is not a
problem as such, but it means that even small fluctuations in laser
intensities will lead to loss of gate fidelity. For the particular
case of the pushing gate, the ac-Stark shifts can be balanced against
the Coulomb energy as discussed in
Ref.~\cite{sasura03:_fast_ql_selec_displace}. Obviously ac-Stark
shifts are common to many gate proposals that uses optical
potentials. An experimentally demanding, but quite general solution is
to compensate the shifts along the lines of
Refs.~\cite{PhysRevLett.90.143602,PhysRevA.66.045401}.  In the present
work, we focus on errors that are more directly related to the motion
of the ions and assume that the push potentials are effectively
non-fluctuating.

.

\subsection{Dimensionless Hamiltonian}
\label{sec:units}

The relative strength of the
Coulomb interaction to the trapping potentials turns out to be
conveniently quantified by
\begin{equation}
  \label{eq:def_epsilon}
  \epsilon
  =
  \frac{\frac{e^2}{\pi\epsilon_0 d}}{m\omega^2d^2}
  =
  \frac{\aosc^2}{d^2}
  \frac{
    \frac{e^2}{\pi\epsilon_0 d}
  }{\hbar\omega}
  ,
\end{equation}
which is the ratio of the energy scale of the Coulomb and trap
potential energies at the equilibrium positions of the ions.  In
Ref.~\cite{sasura03:_fast_ql_selec_displace} it was found that
$\epsilon\ll 1$ is the most promising regime. In oscillator units, the
Hamiltonian for the branch labeled by $\beta$ reads
\begin{equation}
  \label{eq:def_ho_units}
  \begin{split}
    \hH^{(\beta)}_\text{push}(t)
    =&
    \sum_{i=1,2} \frac{1}{2} \left[ \hp_i^2 + \hx_i^2 \right]
    +
    \frac{\epsilon}{4}
    \frac{
      \frac{d^2}{\aosc^2}
    }{
      \left| 1+\frac{\aosc}{d}\left(\hx_2-\hx_1\right)  \right|
    }
    \\
    &-
    \sum_{i=1,2} f^{(\beta)}_i(t)
    \left[ \hx_i + G \hx_i^2
    \right]
    .
  \end{split}
\end{equation}

\subsection{Harmonic approximation}
\label{sec:harm_approx_pushing}

When Eqs.~(\ref{eq:def_eff}) are  specialized to the pushing gate, we
get the following:
\begin{align}
  \label{eq:eff_push}
  E^{(\beta)}(t)
  &
  =
  \frac{1}{4}\epsilon \frac{\aosc}{d}
  \left(
    \frac{\xcl_2^{(\beta)}-\xcl_1^{(\beta)}}
    {1+\frac{\aosc}{d}\left(\xcl^{(\beta)}_2-\xcl_1^{(\beta)}\right)}
  \right)^3
  \\
  F_i^{\beta)}(t)
  &=
  -(-1)^if_i^{(\beta)}(t)\mp
  \frac{1}{4}\epsilon\frac{d}{\aosc}
  \frac{1+3\frac{\aosc}{d}\left(\xcl_2^{(\beta)}-\xcl_1^{(\beta)}\right)}
  {\left[1+\frac{\aosc}{d}\left(\xcl_2^{(\beta)}-\xcl_1^{(\beta)}\right)\right]^3}
  \\
  \label{eq:K_ii}
  \begin{split}
    K_{ii}^{(\beta)}(t)
    &=
    1 + 2 f_i^{(\beta)}(t) G
    \\
    &\phantom{=f_i}+ \frac{1}{2}\epsilon
    \frac{1}
    { \left[ 1 + \frac{\aosc}{d}
        \left(\xcl_2^{(\beta)}-\xcl_1^{(\beta)}\right) \right]^3 }
  \end{split}
  \\
  \label{eq:K_12}
  \begin{split}
    K_{12}^{(\beta)}(t)
    & =
    K_{21}^{(\beta)}(t)
    =
    -\frac{1}{2}\epsilon
    \frac{1}
    { \left[ 1 +
        \frac{\aosc}{d}\left(\xcl_2^{(\beta)}-\xcl_1^{(\beta)}\right)
      \right]^3 }
    .
  \end{split}
\end{align}
When these expressions are inserted into Eqs.~(\ref{eq:eoms_d_b}) and
(\ref{eq:def_h}) we are ready to simulate the gate.

\section{Results of simulation}
\label{sec:results}

\subsection{Choice of parameters}
\label{sec:parameters}
Even with the simplifications we have introduced, there are still a
lot of parameters in the problem. The optimal ``working point'' will
always be dependent on experimental considerations beyond the
simplified model treated here. For a discussion of parameters and
design decisions, see
Ref.~\cite{sasura03:_fast_ql_selec_displace}. For concreteness we have
chosen to focus on a limited set of parameters.  We first of all
assume the individual ion traps to be very well separated and let
$a/d=0.001$ in all calculations. Likewise, we assume a reasonably low
value for $\epsilon$ of $0.04$. Such parameters would result from
e.g.\ $^{40}$Ca ions placed in micro-traps with trapping frequencies
of $\omega\sim 2\pi\times 5 \mathrm{MHz}$ and separated by a distance
of $\sim 7\mu\mathrm{m}$. For a the traveling wave
configuration with beam waist $w$ considered in
Ref.~\cite{sasura03:_fast_ql_selec_displace},
$G=4(a/w)(w/2x_0-2x_0/w)$ where $x_0$ is the initial position of the
ion relative to the beam center. For realistic focusing of the push
beam, $w\sim 1\mu\mathrm{m}$ this suggests to vary the non-linearity
coefficient $G$ between $0$ and $3\times10^{-2}$. As the initial
temporal shape of the push pulse we choose a Gaussian
$f(t)=\xi\exp(-t^2/\tau^2)$, where the amplitude $\xi$ should be
chosen to give a gate phase of $\pi$. A simple estimate (for $G=0$)
suggests that we choose~\cite{calarco01:_entan_micro_traps}
\begin{equation}
  \label{eq:xi}
  \xi^2=\frac{\pi}{\sqrt{\pi/8}\;\epsilon\sqrt{1+\epsilon/2}\;\omega\tau}
  .
\end{equation}
The temporal width of the pulse, $\tau$, should be within an order of magnitude
from the trap period if we  want a fast gate. We will
mainly look at $\tau$ in the range  1 to 10 trap periods, which for the parameters quoted above results a maximum excursion due
to the push in the range from 12$\aosc$ down to 3$\aosc$.

\subsection{Phase errors}
\label{sec:phase_errors}

\begin{figure}[tbp]
  \centering
  \includegraphics{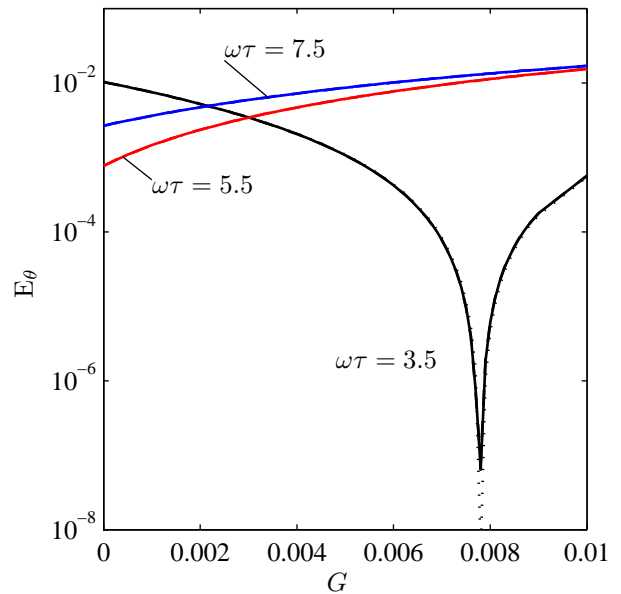}
  \caption{Phase errors with non-uniform pushing forces. The physical
    parameters are: $\epsilon=0.04$, $a/d=0.001$, and $\omega\tau=3.5$
    (black), $5.5$ (red) and $7.5$ (blue). Results for temperatures of
    $T=0.125,1$ and $8\times\hbar\omega/\kB$ are plotted for each
    value of $\tau$ and are indistinguishable from each other.}
  \label{fig:infidel_theta_Gnot0}
\end{figure}

The choice of push amplitude expressed by Eq.~(\ref{eq:xi}) is not
optimal. This can be seen in Fig.~\ref{fig:infidel_theta_Gnot0} where
we plot $\mathrm{E}_\theta$ as a function of $G$. Even for $G=0$ the
gate phase is not exactly $\pi$. For $\omega\tau=3.5$ we see that
a nonzero $G$ can improve the gate phase. This is not surprizing, but
also not very useful as we shall see below that $\mathrm{E}_\theta$ is
in general easy to reduce.
In Fig.~\ref{fig:infidel_theta_Gnot0} results for three different
temperatures are plotted, but the dependence on temperature is
completely negligible.

\subsection{Sloshing errors}
\label{sec:sloshing_errors}

\begin{figure}[tbp]
  \centering
  \includegraphics{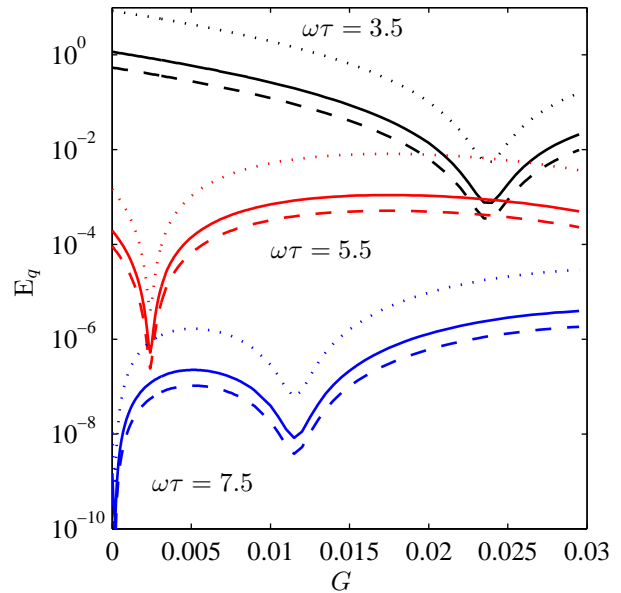}
  \caption{Sloshing errors with non-uniform pushing forces. The
    physical parameters are: $\epsilon=0.04$, $a/d=0.001$, and
    $\omega\tau=3.5$ (black), $5.5$ (red), and $7.5$ (blue).  We plot
    $\mathrm{E}_\vqcl$ as a function of $G$. As explained in the text,
    there exist nonzero values of $G$ where the sloshing is strongly
    suppressed. Results for temperatures of $T=0.125$, 1 and
    $8\times\hbar\omega/\kB$ are plotted as dashed, full and dotted
    lines, respectively.}
  \label{fig:infidel_q_Gnot0}
\end{figure}

\begin{figure}[tbp]
  \centering
  \includegraphics{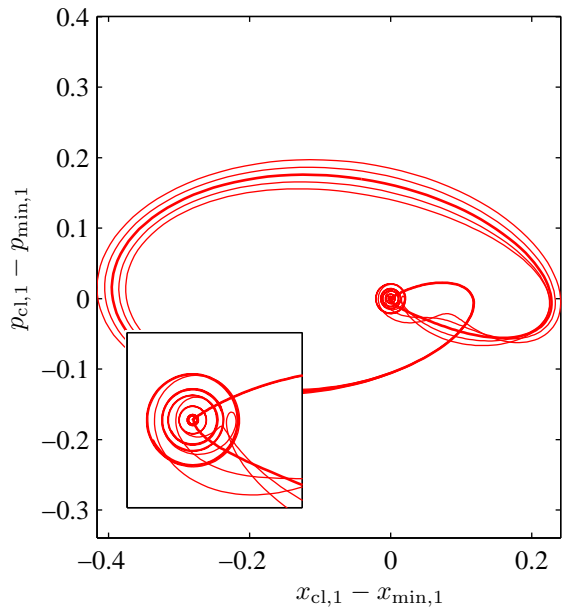}
  \caption{Phase-space trajectories for ion 1 for a push duration of
    $\omega\tau=5.5$. The coordinates are
    relative to the potential minimum in which ion 1 is
    trapped. Perfect adiabatic evolution would correspond to the ion
    simply following this minimum and thus the trajectory would be the
    single point $(0,0)$ in this plot. Since the push is not
    infinitely slow, the ion will first lag behind the moving minimum
    and then oscillate in the moving potential. When the potential
    minimum again approaches its original position, the ion may happen
    to have just the right position and speed in order to end up at
    rest. Whether this is the case depends (for fixed push pulse) on
    $G$: the higher $G$, the more the confinement is increased during
    the push. In the figure, the thick line corresponds to $G=0.002$,
    which is nearly optimal w.r.t.\ returning the ion to rest. The
    thin lines correspond to $G=0,0.001,0.003,0.004$. The lowest $G$
    gives the outermost curve over the main part of the ``loop'' in
    the figure.}\label{fig:phase_space_comb}
\end{figure}

Let us now turn to the errors described by the $\mathrm{E}_\vqcl$ term
in Eq.~(\ref{eq:def_Jobj}), the sloshing errors.
Fig.~\ref{fig:infidel_q_Gnot0}
shows how these errors are strongly dependent on $G$, the strength of
the non-linearity of the pushing potential. A series of minima of
$\mathrm{E}_\vqcl$ as a function $G$ can be seen. The optimal values of
$G$ depends on the chosen duration of the pulse, $\tau$. Each minima
is associated with the ions performing an integer number of
``non-adiabatic oscillations'' during the push pulse. This is
illustrated in Fig.~\ref{fig:phase_space_comb}
where the trajectory of ion 1 with respect to its trap minimum is
plotted for values of $G$ that are below, at and above the one that
leads to the lowest $\mathrm{E}_\vqcl$.

In contrast to $\mathrm{E}_\theta$ above, $\mathrm{E}_\vqcl$ depends
noticeably on whether $T$ is $0.125$, $1$ or
$8\times\hbar\omega/\kB$. Higher temperatures always increase the
sloshing errors and for $\kB T / \hbar\omega \gg 1$ we find that
$\mathrm{E}_\vqcl$ scales as $T$ since $\gamma$ does [see
Eqs.~(\ref{eq:gamma_rel}) and (\ref{eq:gamma_CM})].

Curves for three different values of $\tau$ are plotted in
Fig.~\ref{fig:infidel_q_Gnot0} and it is immediately clear that one can
dramatically decrease sloshing errors by make the gate slower and thus
more adiabatic. The suppression of $\mathrm{E}_\vqcl$ is exponential and
this is thus in general an efficient strategy.

\subsection{Breathing errors}
\label{sec:errors_from_S}

\begin{figure}[tbp]
  \centering
  \includegraphics{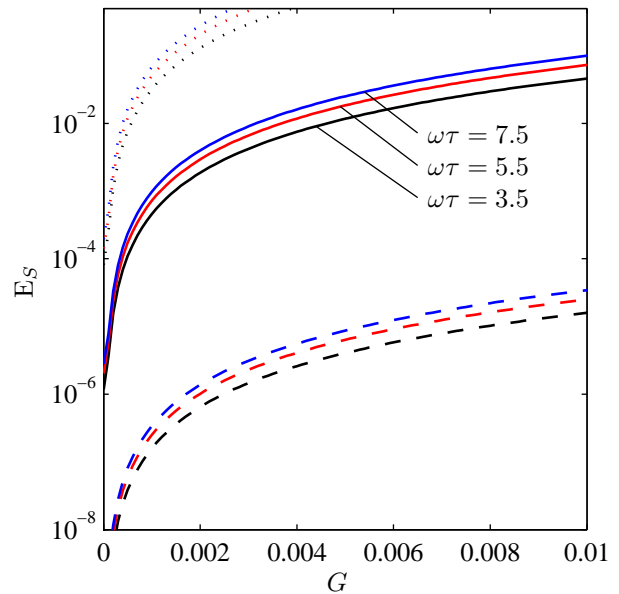}
  \caption{Breathing errors with non-uniform pushing forces. The
    physical parameters are: $\epsilon=0.04$, $a/d=0.001$, and
    $\omega\tau=3.5$ (black), $5.5$ (red), and $7.5$ (blue). Results
    for temperatures of $T=0.125$, 1 and $8\times\hbar\omega/\kB$ are
    plotted as dashed, full and dotted lines, respectively. For all
    three temperatures, larger $\tau$ leads to larger
    $\mathrm{E}_S$. $\mathrm{E}_S$
    is increasing approximately linearly with $G^2\tau$, i.e. for
    fixed $G$, $\mathrm{E}_S$ will be proportional to $\tau$, the
    pulse duration!  This is very different from the behavior of
    $\mathrm{E}_\vqcl$ above, which is rapidly decreased by increasing
    $\tau$ and thereby making the push more adiabatic.}
  \label{fig:infidel_S_Gnot0}
\end{figure}

We now turn to the $\mathrm{E}_S$ term of
Eq.~(\ref{eq:fidelity_split}). These ``breathing'' errors come from
the \emph{different} changes in the effective quadratic Hamiltonian
for the different branches of the evolution. In
Fig.~\ref{fig:infidel_S_Gnot0} we plot $\mathrm{E}_S$ for different
values of $G$ and $\tau$ and for different temperatures of the
external motion. Results for temperatures of $T=$0.125, 1 and 8$\times
\hbar\omega/\kB$ are shown and it is first of all clear that
$\mathrm{E}_S$ depends strongly on $T$.
We also see that $\mathrm{E}_S$ is nearly proportional to $G^2\tau$.
That larger $G$ leads to larger errors is not surprising, but that
larger $\tau$ does is rather counter-intuitive: larger $\tau$ means a
more smooth and thus more adiabatic push. It also means a smaller
amplitude for the push since the ions will have more time to pick up
the gate phase, cf.\ Eq.~(\ref{eq:xi}). Let us discuss the explanation
for this behaviour in more detail.

For exponential suppression of errors to be valid, the evolution
should be well into the adiabatic regime. At first sight, the relevant
timescale is $\omega^{-1}$, the oscillation period of the micro
traps. Since the traps are assumed to be far apart ($a\ll d$) the
parameter $\epsilon$ is small and the normal modes of the system have
periods shifted little from this value: the CM mode is in fact
unaffected by the Coulomb interaction and has frequency $\omega$ while
the relative motion mode oscillates at
$\sqrt{1+\epsilon}\;\omega$. With $\omega\tau \gg 1$ one should
therefore not be able to put excitations into either of these modes.
However, it is perfectly possible to \emph{transfer} excitations
between the modes, as the adiabatic timescale for this process is
$(\sqrt{1+\epsilon}\;\omega - \omega)^{-1} \sim
\epsilon^{-1}\omega^{-1}/2$. Such a transfer will be induced by mixing
of the CM and relative motion during the gate operation. A linear push
potential will not mix the two, but a non-linear one will.

Another effect to remember is that when the instantaneous oscillation
frequencies change during the push, the external motion will pick up
different phases depending on the number of excitation quanta. Like
the transfer of excitations, this effect of course disappears if the
system is cooled to the ground state. A perturbative calculation to
lowest order in $a/d$, $G$ and $\epsilon$ gives the result
\begin{multline}
  \label{eq:perturbative_Es}
  E_S =
  \frac{\pi}{20}G^2\xi^2\omega^2\tau^2\times
  \\
  \times
  \Biggl\{
  \frac{1}
  {\sinh^2\left(\hbar\omega/2\kB T\right)}
  \left[1+\exp\left(-\frac{\epsilon^2\omega^2\tau^2}{8}\right)\right]
  \\
  +2\left[
    \frac{1}{\sinh^2\left(\hbar\omega/2\kB T\right)}+2
  \right]
  \exp\left(-2\omega^2\tau^2\right)
  \Biggr\}
  .
\end{multline}
The prefactor gives the scaling behaviour both in the ``na{\"i}ve''
nonadiabatic limit $\omega \tau < 1$ and in the more relevant
intermediate region $1 < \omega\tau < \epsilon^{-1}$:
\begin{equation}
  \label{eq:dom_G_contrib_to_JS}
  \mathrm{E}_S
  \propto
  \xi^2 G^2 \omega^2\tau^2
  \propto
  \frac{G^2  \omega\tau}{\epsilon},
\end{equation}
In fact, even in the adiabatic limit $\epsilon\omega\tau\gg 1$ this
scaling holds true since one term in Eq.~(\ref{eq:perturbative_Es})
does not contain an exponential damping factor with $\tau^2$. This
unsuppressed term is stemming from the above mentioned effect of
time-varying instantaneous mode frequencies.

From Eq.~(\ref{eq:perturbative_Es}) we can also understand the strong
temperature dependence of the breathing errors. For $\kB
T/\hbar\omega \gg 1$, the breathing errors will scale approximately
like $T^2$.  However, as seen in Figure~\ref{fig:infidel_S_Gnot0}, high temperatures require very low values for
$G$. For low temperatures, note that one term in
Eq.~(\ref{eq:perturbative_Es}) is not suppressed even at $T=0$. This
term stems from changes in the ground state widths of the two
instantaneous normal modes and is adiabatically suppressed when
$\omega\tau\gg 1$. To find the dominant term for very low temperatures
and short pulses one should do a higher order perturbative
calculation.

\section{Optimizing gate performance}
\label{sec:optimal_control}

From the simulations in Sec.~\ref{sec:results} we learn that without
improvement high fidelities require either very low values of $G$ or
cooling the external motion almost to the ground state. In this
section we shall see how a better performance can be achieved by
modifying the temporal shape of the push-pulses.

\subsection{Correcting the phase}
\label{sec:correcting_phase}

\begin{figure}[tbp]
  \centering
  \includegraphics{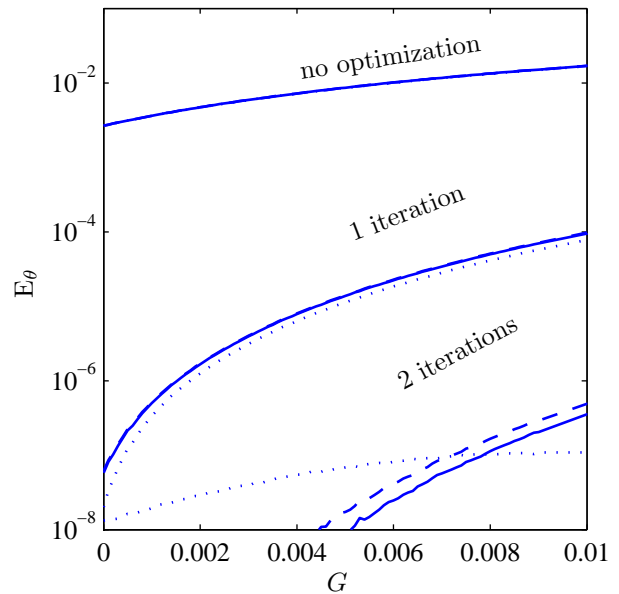}
  \caption{Improving $\mathrm{E}_\theta$ by iteratively adjusting the
    push-amplitude $\xi$. The pulse-duration is $\tau=7.5$ and the
    other parameters are as in the previous figures. Four sets of
    curves are shown, each with results for $T=0.125$ (dashed), 1
    (full), and 8$\times\hbar\omega/\kB$ (dotted). The uppermost set
    (looks like a single curve) is for the un-optimized
    pulse-amplitude and identical to the $\tau=7.5$ curves of
    Fig.~\ref{fig:infidel_theta_Gnot0}. The progressively lower sets
    are for one and two iterations of the amplitude scaling
    described in the text. Note that as the error gets smaller,
    temperature begins to have an effect. This simple adjustment is
    capable of reducing $\mathrm{E}_\theta$ below $10^{-6}$ for all
    the considered $G<0.01$.}
  \label{fig:theta_improve}
\end{figure}

Our first step will be to correct the gate phase by a simple scaling
of the push pulse-shape. From Fig.~\ref{fig:infidel_theta_Gnot0} we
know that typical errors can be well above the percent level. Our
strategy is based on the observation that the simplest estimate of the
gate-phase suggest that it scales as the square of the
push-amplitude, $\xi$. [A general gate-phase replaces the $\pi$ in the
numerator of Eq.~(\ref{eq:xi}).] We therefore divide $\xi$ by the
\emph{square-root} of the ratio of the observed gate-phase and the
ideal gate-phase ($\pi$) and repeat the propagation. In Fig.~\ref{fig:theta_improve}
we show the results of applying this algorithm to the $\tau=7.5$
curves of Fig.~\ref{fig:infidel_theta_Gnot0}. As can be seen,
$\mathrm{E}_\theta$ is rapidly reduced and can be brought below e.g.\
$10^{-6}$ in a very modest number of iterations. For simplicity we
ignore the temperature dependent $\Tr[b\gamma]$ contribution to
$\mathrm{E}_\theta$ when rescaling the pulse. This is the reason for
the $\kB T=8\hbar\omega$ (dotted curves) departing from the $\kB T =
0.125\hbar\omega$ and $\kB T = 1\hbar\omega$ curves especially at low
$G$.

\subsection{Fast gate: eliminating sloshing in $\vxcl$}
\label{sec:faster_gate}

A big advantage of the simple harmonic approximation is that it
becomes feasible to solve the equations of motion many times with
different temporal shapes of $f(t)$ in order to optimize the
performance of the gate. Rather than simple trial-and-error we will
apply the global control algorithm of Krotov, which is guaranteed to improve
the performance at each 
iteration~\cite{krotov95:_global_methods_control,tannor92:_krotov,sklarz02:_loadin_bec_opt_control}. The
relevant equations for our case are given in
Appendix~\ref{sec:krotov}.

In general, it is desirable to complete the gate in as short a time as
possible. This will limit many undesired effects and it will
ultimately enable faster quantum computations. A fast gate, however,
means that the pushing force will deliver a rather abrupt impulse.
This can lead to excitations of the external motion being left after
the completion of the gate, limiting the fidelity. In this section we
show how such ``sloshing'' effects can be avoided by using optimal
control.

\begin{figure}[tbp]
  \centering
  \includegraphics{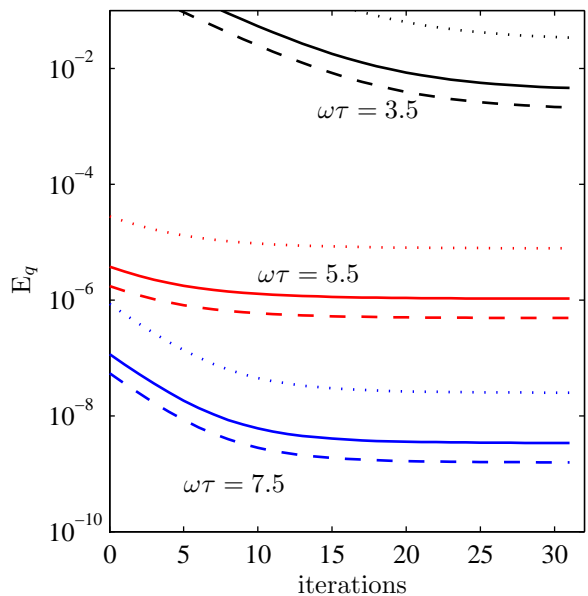}
  \caption{Optimization of pulse-shape to eliminate sloshing. We plot
    $\mathrm{E}_\vqcl$, the contribution of sloshing to the total
    infidelity, as a function of the number of Krotov iterations
    performed. The parameters in this example are: $\epsilon=0.04$,
    $a/d=0.001$, $G=2\times 10^{-5}$. Each curve corresponds to a
    separate value of the pulse duration, $\omega\tau=$3.5, 5.5, 7.5,
    with larger $\tau$ always giving a lower value of
    $\mathrm{E}_\vqcl$.}
  \label{fig:elim_slosh}
\end{figure}

We start from an initial Gaussian temporal shape of the push. The
overall amplitude is first optimized iteratively to get the desired
gate phase as described above. We then run the
Krotov algorithm to get a better shape of the pulse. We  assume a
non-uniform pushing force, $G=2\times 10^{-3}$. The result is
plotted in Fig.~\ref{fig:elim_slosh}.
As can be seen, the influence of sloshing motion can be decreased by a
couple of orders of magnitude in a modest number of iterations.

\begin{figure}[htb]
  \centering
  \includegraphics{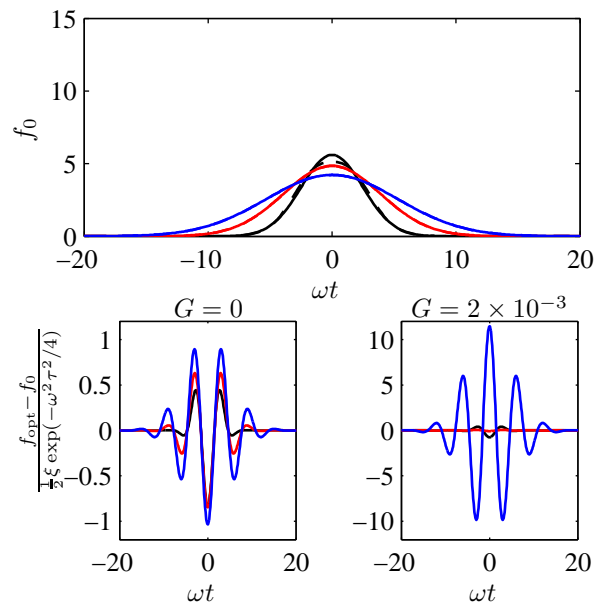}
  \caption{ \label{fig:pulse_shape} Upper panel: pulse shape both
    before (full line) and after (dashed line) Krotov
    optimization. Curves for three different pulse duration are shown,
    $\omega\tau=3.5$, 5.5 and 7.5, other parameters are as in
    Fig.~\ref{fig:elim_slosh}. Only for the shortest pulse is the
    optimized curve distinguishable from the original Gaussian. Lower
    panel: Difference between optimized pulse and initial Gaussian
    pulse (after adjusting overall amplitude to reduce phase
    errors). The left plot is for $G=0$, the right plot is for
    $G=2\times 10^{-3}$. Each curve has been normalized to the
    prediction of a $G=0$
    perturbative calculation, which is seen to describe well the $G=0$
    case  as all curves have a maximal excursion of approximately -1,
    while the $G\neq 0$ case is only qualitatively similar.}  
\end{figure}

To investigate the physical mechanism behind the reduction of the
sloshing error, we plot in the lower panel of
Fig.~\ref{fig:pulse_shape}
the difference between the optimized pulse $f_\mathrm{opt}(t)$ and the
original Gaussian pulse $f_0(t)=\xi\exp(-t^2/\tau^2)$. This difference
looks a lot like a simple cosine-wave with a period close to
$2\pi\omega^{-1}$ multiplied by a Gaussian of the same width as
$f_0$. Thus the optimized pulse is approximately of the form
$f_\mathrm{opt}(t)\sim f_0(t)+A\cos(\omega t)\exp(-t^2/\tau^2)$. An
approximative calculation of the sloshing excitation for the simplest
$G=0$ case reveals that for such a pulse, the \emph{non-resonant}
contribution of the bare Gaussian pulse is cancelled by a
\emph{resonant} contribution from the cosine-modulated pulse. Since
the resonant response is much stronger, only a small, negative $A$ is
needed for this cancellation. More precisely, the optimal $A$ from
first order pertubation theory is given by
$-\frac{1}{2}\xi\exp(-\omega^2\tau^2/4)$ and Fig.~\ref{fig:pulse_shape}
shows that this is also what the Krotov algorithm converges to for $G=0$. The
``strategy'' of the Krotov algorithm in this case seems therefore to
be well understood. For $G\neq 0$, it is more difficult to predict the
value of $A$, but nonetheless the Krotov algorithm seems to be highly
efficient.

\subsection{Minimizing breathing errors}
\label{sec:breathing_errors_consider}

We now know that phase errors and sloshing errors can be controlled
and we turn to the breathing errors of Fig.~\ref{fig:infidel_S_Gnot0}.
Without optimization, these errors put rather stringent limits on the
parameters.  In order to keep $\mathrm{E}_S$ at an acceptable level,
either very low temperature or very small $G$ is required. For very
low temperature $\kB T
=0.125\hbar\omega$, we need just $G\lesssim 10^{-2}$ to get
$\mathrm{E}_S$ below $10^{-4}$, but if we assume a more modest cooling
to $\kB T =1\hbar\omega$ the same error-level require $G\lesssim
2\times 10^{-4}$.  Note that even at $G=0$, breathing errors persist
and that for $\kB T= 8 \hbar\omega$ they never get below the 10$^{-5}$
level. These errors stem from the high order terms in the Coulomb
potential which have been ignored in Eq.~(\ref{eq:perturbative_Es}).

As described in
Sec.~\ref{sec:errors_from_S}, breathing errors cannot be eliminated by
simply increasing the push duration $\tau$: First of all the adiabatic
time-scale is $\sim \epsilon^{-1}\omega^{-1}/2$ which will mean a slow
gate and secondly even in that limit errors from the change in normal
modes frequencies remain and even increase,
cf.\ discussion below Eq.~(\ref{eq:dom_G_contrib_to_JS}).
It turns out that a simple application of the Krotov algorithm is also
not very efficient in reducing the breathing errors. A partial
explanation for this can be found from the perturbative calculation
leading to Eq.~(\ref{eq:perturbative_Es}) and the analysis of sloshing
error-reduction above: since the adiabatic time-scale for the
breathing errors is long, the Gaussian pulses we consider are not
adiabatic w.r.t.\ breathing errors and thus the admixture
of a small resonant component in the push-pulse will not be enough to
get the cancellation we found in the case of sloshing errors. In fact,
the amplitude of the cosine modulation should be comparable to the
total amplitude for the relatively short pulses considered. There is
nothing to be gained from a small amplitude modulation and thus the
``linear'' version of the Krotov algorithm we apply (see
Appendix~\ref{sec:krotov}) will not work.

\begin{figure}[tb]
  \centering
  \includegraphics{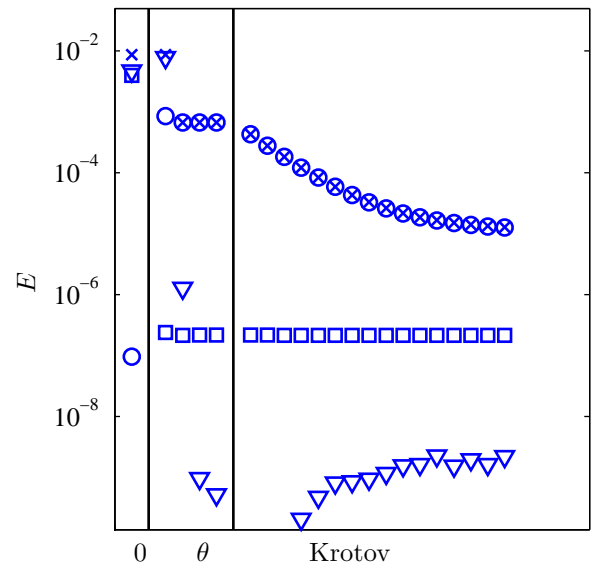}
  \caption{Combined strategy for reducing infidelity for
    $\epsilon=0.04$ and $G=2\times 10^{-3}$. The three contributions
    $E_\theta$, $E_q$, and $E_S$ are labeled by $\triangledown$,
    $\circ$, and $\square$, respectively. Their sum is labeled by
    $\times$. The leftmost column are the results for a simple
    Gaussian with $\omega\tau=7.5$. The breathing errors dominate. In
    the next column, breathing errors have been reduced (but sloshing
    increased) by using a cosine-modulated pulse based on
    Eq.~(\ref{eq:perturbative_Es}). The overall amplitude is then
    iteratively optimized to reduce phase errors as described in
    Sec.~\ref{sec:correcting_phase}. As can be seen, 3 iterations are
    more than sufficient to render $E_\theta$ completely
    insignificant. The dominating error type is now sloshing and in
    the third column, the Krotov algorithm is applied to finally
    reduced the total infidelity below $10^{-4}$.}
  \label{fig:opt_pulse_hand_Krotov}
\end{figure}

In fact, in order to cancel out the contribution to the breathing
errors due to mode frequency changes, sign changes in the push
amplitudes are required during the pulse. This is beyond the simplest
physical implementations where the push amplitude is proportional to
some laser intensity. In principle it is possible to play with
detunings to implement the sign changes, and this will in fact give
many of the advantages of the $\pi$-pulse method, see
Ref.~\cite{sasura03:_fast_ql_selec_displace}.  Allowing negative push
amplitudes and putting ``by hand'' an optimized $\cos(\epsilon\omega
t/2)$ modulated contribution, we have been able to e.g.\ reduce $E_S$
below $10^{-6}$ for $\omega\tau=7.5$, $G=2\times10^{-3}$, and
$T=1\times\hbar\omega/\kB$. Compared to the results reported in
Fig.~\ref{fig:infidel_S_Gnot0}, this is a reduction by more than 3
orders of magnitude. Unfortunately, the strongly modified pulse now
gives rise to large sloshing errors. To obtain an overall safisfactory
fidelity we use a combined strategy: we first put the
breathing error reduction by hand, then iteratively reduce phase
errors and finally use the Krotov algorithm to reduce the sloshing
errors. In Fig.~\ref{fig:opt_pulse_hand_Krotov}
we show results of this strategy starting from $\omega\tau=7.5$.

\section{Conclusion}
\label{sec:conclusion}
In this paper we have shown how a time-dependent, quadratic
approximation to the Hamiltonian can be a useful tool when analyzing
quantum gates based on conditional external dynamics. The resulting
equations are much more manageable than the original two-body
Schr\"{o}dinger equations. This is especially true if one were to
include more spatial dimensions than the one considered here: A full
time-dependent, three dimensional, two-body wavefunction calculation
is an extremely demanding numerical task, whereas the corresponding
quadratic approximation will be much more manageable.

We used the developed method to show how to improve on a na\"{i}ve
design of the pushing gate. This was done including a non-uniform
contribution to the pushing force. An important lesson of our analysis
and simulations, is to pay attention to changes in the symmetry of the
Hamiltonian during the gate operation. In the present case, a
nonlinear push potential invalidates the separation of the dynamics
into CM and relative motion. This opens up another type of
non-adiabaticity, namely transfer of excitations between the two
normal modes. The adiabaticity parameter for this type of error is
$\epsilon\omega\tau$ and for small $\epsilon$, adiabaticity will
require gate times much larger than the charateristic time of the
micro-traps. An efficient counter-measure is to decrease temperature
so that there are in fact no excitations to transfer between
modes. Failing that, one should increase $\epsilon$ as much as
possible and, somewhat counter-intuitively, do the gate as fast
possible. Sloshing errors puts a lower limit on the gate-time, but as
we show an optimized choice of the temporal shape of the push can
dramatically reduce this problem.

By analyzing the way that the Krotov optimized pulse reduces sloshing
errors, we identified the basic mechanism as a destructive
interference between the non-resonant, non-adiabatic contribution from
the finite push pulse duration and a resonant contribution from a small
amplitude superposed oscillation of the push force. Generalizing this
idea to deal with breathing errors, we were able to reduce them by
several orders of magnitude. However, since breathing errors are not significantly adiabatically
suppressed for the considered pulse
durations, the destructive interference required
sign-changes in the push amplitude, which introduce experimental
complications. The suppression of breathing errors also came at the
price of increased sloshing errors, but we showed that the Krotov
algorithm once again was able to improve the pulse shape.

One may ask to what extent is the final pulse shape optimal for the
given overall gate-time? This is an interesting question in general
and in this study we saw examples both where the Krotov algorithm
seemed to exhaust the potential in its ``strategy'' (the $G=0$ case in
Fig.~\ref{fig:pulse_shape}) and where it was not able to find an
optimization. In the latter case we could improve the pulse by hand (eliminate the breathing errors by
destructive interference). The problem of optimality is related to
the question of a \emph{quantum speed limit} (QSL)~\cite{PhysRevA.67.052109} and this connection
has been studied in Ref.~\cite{caneva09:_opt_control_quan_speed_limit}. Note, however, that in our case the
hamiltonian is time-dependent and what we want is in fact to leave the
external motion unaffected after the pulse. It would be interesting to
investigate such a general adiabaticity problem along the same lines
as the work on the QSL.

\acknowledgements
This research was supported in part by the National Science Foundation under Grant No. PHY05-51164, and in part by the EC projects AQUTE and EMALI.

\appendix

\section{Evolution under quadratic Hamiltonian}
\label{sec:evol_quad_ham}
In this section we show that a second order Hamiltonian leads to an
evolution operator that can be written like $\hU(t)$
Eq.~(\ref{eq:evol_operator}). For alternative parameterizations, see
Refs.\cite{gilmore87:_group_theo_semiclas_dyn_single_mode,gilmore89:_group_theo_semiclas_dyn_multimode}. We
will do a direct calculation showing that $\hU(t)$ fulfills the
Schr\"{o}dinger equation
\begin{equation}
  \label{eq:schrodinger}
  i\frac{\partial}{\partial t} \hU = \hH_\mathrm{so}\hU.
\end{equation}
The demanding part of the the calculation involves differentiating
expontials of time dependent operators. A useful formular can be found
in e.g.~\cite{wilcox67:_expon_operat_param_differ_quant_physic} and
involves integration over an auxillary variable $\wilcox$. It results in
\begin{equation}
  \label{eq:dDdt}
  \begin{split}
  i\frac{\partial}{\partial t} \hD_\vc
  = &
  \int_0^1
  \hD_{\wilcox\vc} \;
  \dot{\vc}^\mathrm{T} J \hvq \;
  \hD_{\wilcox\vc}^\dagger \; d\wilcox \; \hD_\vc
  \\
  = &
  \int_0^1
  \dot{\vc}^\mathrm{T} J \left( \hvq - \wilcox \vc \right) \;
  d\wilcox \; \hD_\vc
  \\
  = &
  \dot{\vc}^\mathrm{T} J \left( \hvq - \frac{1}{2} \vc \right)
  \hD_\vc
\end{split}
\end{equation}
for the displacement operator and
\begin{equation}
  \label{eq:dWdt}
  \begin{split}
  i\frac{\partial}{\partial t} \hW_b
  = & \int_0^1 \hW_{\wilcox b} \; \frac{1}{2}\hvq^\mathrm{T} \dot{b}
  \hvq  \hW_{\wilcox b}^\dagger \; d\wilcox \; \hW_b
  \\
  = &
  \frac{1}{2}\int_0^1
  \left(e^{-\wilcox Jb} \hvq\right)^\mathrm{T}
  \dot{b}
  \left(e^{-\wilcox Jb}\hvq\right) \; d\wilcox \; \hW_b
  \\
  = &
  \frac{1}{2}\hvq^\mathrm{T} J^\mathrm{T} \int_0^1
  e^{\wilcox J b}
  J \dot{b}
  e^{-\wilcox Jb} \; d\wilcox \; \hvq \hW_b
  \\
  =&
  \frac{1}{2}\hvq^\mathrm{T} J^\mathrm{T}
  \left(\frac{\partial}{\partial t}
  e^{J b}
  \right) e^{-Jb} \; \hvq
  \; \hW_b
\end{split}
\end{equation}
for the squeezing operator. It is now easy to show that the equations
of motion (\ref{eq:eoms_d_b}) for $\phi$, $\vc$ and $S=\exp(Jb)$ leads
to $\hU=\exp(-i\phi)\hD_\vc\hW_b$ fulfilling
Eq.~(\ref{eq:schrodinger})~\endnote{There are two caveats regarding
  the equation of motion for $S$. First of all, the translation from
  $S$ to $b$ and thus $\hW_b$ is not one-to-one. Secondly, the
  equation of motion for $S$ cannot necessarily be fulfilled by an $S$
  in the form $\exp(Jb)$ with a differentiable $b(t)$. Both problems
  are however eliminated when the $S$ in the four
  branches never deviate much from eachother.}.

\section{Fidelity for Gaussian evolutions}
\label{sec:fidelity}
In this section we derive expressions for the fidelity as a function
of the variables used to characterize the evolution in the harmonic
approximation, $\phi_\beta$, $\vqcl_\beta$, and $S_\beta$,
$\beta=00,01,10,11$.

If we assume that the initial state of the system is a product of an
internal state density matrix and an external state density matrix,
$\rho\otimes\sigma$, we get the following state for the internal
degrees of freedom after the application of $\hU^\text{tot}$ of
Eq.~(\ref{eq:con_dyn_U}):
\begin{equation}
\label{eq:hadamard_form}
\begin{split}
  \rho'
  =&
  \Tr_\text{ex}
  \left[
    \hU^\text{tot}
    \rho\otimes\sigma
    \left(\hU^\text{tot}\right)^\dagger
  \right]
  \\
  =&
  \sum_{\alpha\beta}
  \dyad{\alpha}{\beta}
  \;
  [\rho\circ R]_{\alpha\beta}
  .
\end{split}
\end{equation}
Here ``$\circ$'' denotes the element-wise matrix product (the Hadamard
product) and $R$ is the matrix given by:
\begin{equation}
  \label{eq:def_R}
  [R]_{\alpha\beta}
  =
  \Tr
  \left[
    \hU^\text{ex}_{\alpha}
    \sigma
    \left(\hU^\text{ex}_{\beta}\right)^\dagger
  \right]
\end{equation}
It is easy to see that $R$ er Hermitian and that all its diagonal
elements are 1. In particular, $\Tr R = 4$. Slightly less obvious is
it that $R$ is positive semi-definite: Let $c \in \mathbb{C}^4$. Then:
\begin{equation}
  \begin{split}
  c^\dagger R c
  &=
  \Tr
  \left[ \bigl\{ \sum c_{\alpha}^* \hU^\text{ex}_{\alpha}\bigr\}
    \sigma
    \bigl\{\sum c_{\beta} \left(\hU^\text{ex}_{\beta}\right)^\dagger\bigr\}
  \right]
  \\
  &=
  \Tr
  \left[
    \bigl\{\sum c_{\beta}^* \hU^\text{ex}_{\beta}\bigr\}^\dagger
    \bigl\{\sum c_{\alpha}^* \hU^\text{ex}_{\alpha}\bigr\}
    \sigma
  \right]
  \ge
  0
\end{split}
\end{equation}
where we have used the cyclic property of the trace and the fact that
the trace of a product of positive semi-definite operators is
non-negative.

The element-wise product form in Eq.~(\ref{eq:hadamard_form}) is
perhaps not the most illuminating. If we diagonalize $R=\sum_h
 w_h w_h^\dagger$, we get instead a Krauss operator sum form:
\begin{equation}
  \label{eq:diag_form}
  \rho'
  =
  \sum_{h=1}^{4} K_h \rho K_h^\dagger
\end{equation}
with $K_h=\mathrm{diag}(w_h)$.

There are different ways to define the fidelity of the gate. The one
used in Refs.~\cite{sasura03:_fast_ql_selec_displace}
and~\cite{calarco01:_entan_micro_traps} is as the minimum fidelity of
the obtained final state w.r.t.\ the wanted final state when the input
state is varied. This means that
\begin{equation}
  \label{eq:def_min_fidelity}
  F_{\min} =\min_\psi \bra{\psi} U_0^\dagger \rho_\psi' U_0 \ket{\psi}
\end{equation}
with $\rho_\psi=\dyad{\psi}{\psi}$ and $U_0$ the gate operator we aim
for. In the case of a phase gate, $U_0$ is diagonal in the logical
state basis:
\begin{equation}
  \label{eq:def_U0}
  \hU_0
  =
  \sum_{\beta}
  \dyad{\beta}{\beta}e^{i\theta_{\beta}}
\end{equation}
and we get the simpler minimization problem:
\begin{equation}
  \label{eq:simple_F}
  F_{\min}=\min_{\{p_i\}} p^\mathrm{T} \tilde{R}  p
\end{equation}
where the $p\in\mathbb{R}_+^4$ and $\sum p_i = 1$. The matrix
$\tilde{R}= U_0 R U_0^\dagger$ is Hermitian, but since $p$ is confined
to be real, only its real symmetric part contributes.  The
minimization in Eq.~(\ref{eq:simple_F}) is a so-called quadratic
programming problem and very efficient numerical methods for its
solution exist. Given $\tilde{R}$ it is therefore simple to calculate
$F_{\min}$ on the computer. However, a more direct evaluation is
possible if fidelity is instead defined as an \emph{average} over
input states as
\begin{equation}
  \label{eq:def_aver_fidelity}
  F_\mathrm{avg}
  =
  \int_{S^{2n-1}}
  \bra{\psi} U_0^\dagger \rho_\psi' U_0 \ket{\psi}
  \;dV
  .
\end{equation}
Here $S^{2n-1}$ denotes the normalized states (unit sphere) in
$\mathbb{C}^n$ and the volume element $dV$ is such that
$\int_{S^{2n-1}} dV =1$. For $F_\mathrm{avg}$ a compact formula
exist~\cite{pedersen07:_fidel_q_operations} and using it in the
present case leads to
\begin{equation}
  \label{eq:F_of_R}
  F_\mathrm{avg}
  =
  \frac{1}{4(4+1)}
  \left(
    \Tr \tilde{R} +\sum_{\alpha\beta}\tilde{R}_{\alpha\beta}
  \right)
\end{equation}
or when using the properties of $\tilde{R}$:
\begin{equation}
  \label{eq:infidel_of_R}
  1-F_\mathrm{avg}
  =
  \frac{1}{10}
  \sum_{\alpha< \beta}
  \left(
    1-\mathrm{Re}\left[\tilde{R}_{\alpha\beta}\right]
  \right)
  .
\end{equation}

\subsection{General small errors}
\label{sec:gen_small_errors}

Typically, we will be mostly interested in situations where the four
logical states leads to almost identical evolutions for the external
states. It is then useful to write
\begin{equation}
  \label{eq:def_D}
  \begin{split}
    \hU^\text{ex}_{\beta}
    =&
    \exp(i \hD_{\beta}) \hU^\text{ex}_\mathrm{com}
    \\
    =&
    \exp\left(
      i \langle \hD_{\beta} \rangle
    \right)
    \exp(i \Delta \hD_{\beta}) \hU^\text{ex}_\mathrm{com}
    ,
  \end{split}
\end{equation}
where $\langle \hD\rangle = \Tr\left[\hD
  \hU^\text{ex}_\mathrm{com}\sigma_\mathrm{ex}(\hU_\mathrm{com}^\text{ex})^\dagger\right]$
and $\Delta \hD = \hD - \langle \hD \rangle$. Calculating $\tilde{R}$ to
second order in the $\Delta \hD$'s, we get:
\begin{equation}
  \label{eq:R_2nd_order}
  \begin{split}
    \tilde{R}_{\alpha\beta}
    =&
    \exp\left(
      i \Delta\theta_\alpha 
    \right)
    \\
    &\times
    \left\{
      1-
      \frac{1}{2}
      \left\langle
          \Delta \hD_{\alpha}^2
          +\Delta \hD_{\beta}^2
          -2\Delta \hD_{\alpha}\Delta \hD_{\beta}
      \right\rangle
    \right\}
    \\
    &\phantom{=}\times
    \exp\left(
      -i \Delta\theta_\beta 
    \right)
  .
\end{split}
\end{equation}
This form is useful, as it separates the infidelity into systematic
phase-errors (the $\exp(\pm i\Delta\theta_\alpha)$ factors) and
``decoherence'' (factor in curly brackets). The phase errors $\Delta
\theta_{\beta}=\langle \hD_{\beta}\rangle-\theta_{\beta}$ can be made
small by tuning the average of laser powers etc.\ and this can usually
be done very well. The challenge will therefore most often be to
suppress the fluctuations, i.e., the terms in curly brackets in
Eq.~(\ref{eq:R_2nd_order}).

Assuming also the $\Delta \theta$'s to be small,
Eq.~(\ref{eq:infidel_of_R}) becomes:
\begin{equation}
  \label{eq:infidel_2_order}
  1-F_\mathrm{avg}
  =
  \frac{1}{20}\sum_{\alpha < \beta}
  \left(
    \left[ \Delta \theta_\alpha- \Delta \theta_\beta \right]^2
    +
    \left\langle \left(\Delta \hD_\alpha- \Delta \hD_\beta \right)^2\right\rangle
  \right)
  .
\end{equation}
At first sight, this form might seem dubious since only
\emph{differences} in the $\Delta \theta$'s and $\Delta \hD$'s
enter. However, one should remember that any \emph{common} evolution
on the four branches can be absorbed into $\hU^\text{ex}_\mathrm{com}$ in
Eq.~(\ref{eq:def_D}). This emphasizes that for the implementation of a
single gate on the logical state, the external motion must not
necessarily be returned to its initial state as long as the final
state is common to all logical input states. Typically, the further
requirement that energy is not pumped into the external degrees of
freedom by repeated application of the gate must be made. In the
particular case of the pushing gate, this requirement is in fact
already hidden in Eq.~(\ref{eq:infidel_2_order}) since the $\beta=00$
branch contains no pushing. In other cases, one could apply cooling to
the external state between gate operations.

\subsection{The non-local part of the phase}
\label{sec:non-local_phase}
We are seeking to implement the phase-gate (\ref{eq:def_U0}). In many
cases, the $\theta_\beta$'s are not so important individually since
single-particle operations are easy to perform and only the truely
non-local phase is interesting. Assuming that perfect single-particle
phase changes can be implemented \emph{on average}, it is
straightforward to show that one should replace
$\sum_{\alpha<\beta}[\Delta \theta_\alpha -\Delta \theta_\beta]^2$ by
\begin{equation}
  \label{eq:non-local_phase}
  \left[
    \Delta \theta_{00} -\Delta \theta_{01}
    -\Delta \theta_{10} + \Delta \theta_{11}
  \right]^2
\end{equation}
in Eq.~(\ref{eq:infidel_2_order}).

This simplified view of single-particle phase-changes should of course
be revisited in a more complete analysis of any given proposal for
quantum-computing. In the present work we use
the replacement (\ref{eq:non-local_phase}) throughout, but let us
emphasize that \emph{fluctuations} in the single-particle
phaserotations are more naturally incorporated in the $\Delta \hD$
terms of Eq.~(\ref{eq:infidel_2_order}) than in the $\Delta \theta$
terms: One simply model the fluctations as a consequence of some
fluctuating parameter which can be included in $\sigma_\mathrm{ex}$.

\subsection{The Gaussian case }
\label{sec:Gaussian_case}
For Gaussian evolutions like (\ref{eq:evol_operator}) and a Gaussian
(e.g. thermal) external state $\sigma$ with \emph{covariance} matrix
\begin{equation}
  \label{eq:def_covar_mat}
  \begin{split}
    \gamma_{ij}
    =&
    \frac{1}{2} \langle q_i q_j + q_j q_i \rangle
    - \langle q_i \rangle \langle q_j \rangle
    \\
    =&
    \mathrm{Re} \Tr \left[
      q_i q_j \sigma
    \right]
    -\Tr \left[q_i\sigma\right] \Tr \left[q_j\sigma\right]
  \end{split}
\end{equation}
and vanishing means
\begin{equation}
  \label{eq:vanish_mean}
  \langle q_i \rangle = \Tr\left[q_i\sigma\right] = 0
\end{equation}
one gets phase contributions
\begin{equation}
  \label{eq:phase_contrib}
  \langle \hD_\beta \rangle = -\phi_\beta - \Tr \left[ b_\beta \gamma \right]
\end{equation}
and decoherence terms
\begin{multline}
  \label{eq:exp_D_gauss}
  \left\langle\left( \Delta D_\alpha -\Delta D_\beta\right)^2 \right\rangle
  =
  (c_\alpha - c_\beta)^\mathrm{T} J \gamma J^\mathrm{T}(c_\alpha - c_\beta)
  \\
  +\frac{1}{2} \Tr \left[(b_\alpha - b_\beta) \gamma
    (b_\alpha - b_\beta) \gamma\right]
  \\
  +\frac{1}{8} \Tr \left[(b_\alpha - b_\beta) J(b_\alpha - b_\beta) J \right]
\end{multline}

In general, for a harmonic oscillator in thermal equilibrium at temperature $T$,
the covariance matrix is given by
\begin{equation}
  \label{eq:gamma_single_thermal}
  \gamma_\mathrm{thermal}
  =
   \frac{1}{2}\frac{1}{\tanh \frac{\hbar\omega}{2k_BT}}
  \begin{bmatrix}
    \frac{\hbar}{m \omega}
    & 0
    \\
    0 &
    \hbar m\omega
  \end{bmatrix}
\end{equation}
where $k_B$ is the Boltzmann constant. In the present case, the CM and
the relative motion are separately in thermal equilibrium and for the
corresponding dimensionless position and momentum operators
[$x_\mathrm{CM}=(x_1+x_2)/2$ etc.], we get:
\begin{equation}
  \label{eq:gamma_GS}
  \gamma
  =
  \gamma_\mathrm{CM}
  \oplus
  \gamma_\mathrm{rel}
\end{equation}
with
\begin{equation}
  \label{eq:gamma_rel}
  \gamma_\mathrm{rel}
  =
  \frac{1}{2}\frac{1}{\tanh\frac{(1+\epsilon)^{1/2}\hbar\omega}{2k_BT}}
  \begin{bmatrix}
    \frac{2}{(1+\epsilon)^{1/2}} & 0 \\
    0 & \frac{(1+\epsilon)^{1/2}}{2}
  \end{bmatrix}
\end{equation}
and
\begin{equation}
  \label{eq:gamma_CM}
  \gamma_\mathrm{CM}
  =
  \frac{1}{2}\frac{1}{\tanh\frac{\hbar\omega}{2k_BT}}
  \begin{bmatrix}
    \frac{1}{2}& 0\\
    0 & 2
  \end{bmatrix}
  .
\end{equation}
In the limit $\epsilon \ll 1$, a we have approximately
$\gamma\propto \Id_4$ if we use the set of individual-ion operators
$(x_1,x_2,p_1,p_2)$.

\section{The Krotov Algorithm}
\label{sec:krotov}
Optimizing the temporal shape of the push pulse is done using the
Krotov
algorithm~\cite{krotov95:_global_methods_control}. For an introduction
to the method, see e.g.~Ref.~\cite{sklarz02:_loadin_bec_opt_control}.

\subsection{Auxillary variables}
\label{sec:aux_var}
The key ingredient in this approach is a function
$\Phi(t,\{\phi_\beta,\vqcl_\beta,S_\beta\}_{\beta=00,01,10,11})$ which
allow us to translate the \emph{global} goal of improving the final
$\hU^\text{tot}(T)$ to a \emph{local} problem of choosing a better
$f(t)$ for each $t$.  Constructing $\Phi$ is in general
very difficult, but it is relatively simple to get a linear
approximation to it. The coefficients in this approximation will
constitute a set of auxillary variables. For each branch, the
equations of motion for the auxillary variables $\cphi_\beta$,
$\vcqcl_\beta$ and $\cS_\beta$ are determined by the requirement that
they are conjugate to the physical variables $\phi_\beta$,
$\vqcl_\beta$, and $S_\beta$, respectively.

Let us focus on a single branch and suppres the $\beta$ index like in
Sec.~\ref{sec:harmonic_approx}. We then need to construct
$\kH(t;\phi,\vqcl,S;\cphi,\vcqcl,\cS)$ such that
Eqs.~(\ref{eq:eoms_d_b}) can be written
\begin{align}
  \label{eq:eoms_as_canon}
  \frac{\partial}{\partial t} \phi &= \frac{\partial}{\partial \cphi}\kH
  \\
  \frac{\partial}{\partial t} \qcl_i &= \frac{\partial}{\partial \cqcl_i}\kH
  \\
  \frac{\partial}{\partial t} S_{ij} &= \frac{\partial}{\partial \cS_{ij}}\kH
  .
\end{align}
This leads simply to
\begin{equation}
  \label{eq:krotov_H}
  \begin{split}
  \kH(t;\phi,\vqcl,S;\cphi,\vcqcl,\cS)
  &=
  \cphi\Bigl[E(t,\vxcl)-\frac{1}{2}\vF^\mathrm{T}(t,\vxcl)\Bigr]
  \\
  &\phantom{=}
  +\vcqcl^\mathrm{T}\Bigl[Jh(t,\vxcl)\vqcl-J\vF(t,\vxcl)\Bigr]
  \\
  &\phantom{=}
  +\Tr\Bigl[\cS^\mathrm{T} J h(t,\vxcl)S \Bigr]
\end{split}
\end{equation}
Then the equations of motion for the auxillary variables become
\begin{align}
  \label{eq:eoms_auxphi}
  \frac{\partial}{\partial t} \cphi
  &=
  -\frac{\partial}{\partial\phi}\kH
  = 0
  \\
  \label{eq:eoms_auxq}
  \frac{\partial}{\partial t} \vcqcl
  &=
  -\nabla_\vqcl\kH
  \\
  \label{eq:eoms_auxS}
  \frac{\partial}{\partial t } \cS
  &=
  \nabla_S \kH
  =
  h(t,\vxcl)J\cS.
\end{align}
The equation of motion for $\vcqcl$ is rather involved since $\kH$
depends on $\vxcl$ in a complicated manner through $h$, $F$ and
$E$. It can be rewritten as two coupled time-dependent, forced
harmonic oscillators. Note on the other hand that $\cphi$ is
time-independent and that $J\cS$ solves the same equation as $S$.

\subsection{Objective function}
\label{sec:obj_function}

Our ultimate goal is to improve the fidelity of the gate. However, it
is somewhat impractical to apply this as the objective in the Krotov
algorithm: Calculating the fidelity is only simple for small errors
and in general it depends on e.g.\ the temperature of the external
motion. Instead we shall work with a simpler function of the variables
$\phi$, $\vqcl$ and $S$ for the four branches. The reduction of this
\emph{objective function} should tend to increase the fidelity of the
gate. Based on the fact that in the pushing gate, the branch
$\beta=00$ is not subject to any time-dependent forces, we choose the
following:
\begin{equation}
  \label{eq:def_Jobj}
  \begin{split}
    \Jobj
    &
    \left(\{\phi_\beta,\vqcl_\beta,S_\beta\}_{\beta=00,01,10,11}\right)
    = \Jobj_\phi+\Jobj_q+\Jobj_S
    \\
    &\phantom{\beta}=
    \frac{1}{2}\Bigl[\phi_{00}-\phi_{01}-\phi_{10}+\phi_{11}-\pi \Bigr]^2
    \\
    &\phantom{\beta=}+
    \frac{1}{2}\sum_\beta \left[
      \Bigl(\vx_\beta-\vx_{00}\Bigr)^2
      +\Bigl(\vp_\beta-\vp_{00}\Bigr)^2
    \right]
    \\
    &\phantom{\beta=}+
    \frac{1}{2}\sum_\beta
    \Tr\left[
      \Bigl(S_\beta-S_{00}\Bigr)^\mathrm{T}\Bigl(S_\beta-S_{00}\Bigr)
    \right]
    .
  \end{split}
\end{equation}
The term with $\phi$'s aim to ensure the correct phase in the
phasegate, while the other terms aim at identical evolution for the
external motin in the four branches. In the limit $\epsilon \ll 1$
Eq.~(\ref{eq:infidel_2_order}) formally justifies the use of our
chosen objective function, given the extra proviso that we are only
interested in the non-local part of the phase.

\subsection{Terminal conditions}
\label{sec:terminal_conditions}
The objective function supplements the auxillary-variable
equations of motion (\ref{eq:eoms_auxphi}--\ref{eq:eoms_auxS}) with
the following \emph{terminal conditions}, i.e. boundary conditions at
$t=T$:
\begin{align}
  \label{eq:auxphi_terminal}
  \cphi_\beta&(T)
  =
  \left. -\frac{\partial}{\partial \phi_\beta} \Jobj\right|_T
  \nonumber \\
  &=
  -(-1)^\beta
  \left.
    \bigl(
    \phi_{00}-\phi_{01}-\phi_{10}+\phi_{11}-\pi
    \bigr)
  \right|_T
  \\
  \label{eq:auxx_terminal}
  \vcxcl_\beta&(T)
  =
  \left.
    -\nabla_{\vxcl_\beta} \Jobj
  \right|_T
  \nonumber \\
  &=
  \begin{cases}
    \left.
      -\Bigl(3\vxcl_{00}-\vxcl_{01}-\vxcl_{10}-\vxcl_{11}\Bigr)
    \right|_T,
    &
    \beta=00
    \\
    \left.
      -\Bigl(\vxcl_\beta-\vxcl_{00}\Bigr)
    \right|_T,
    &
    \beta\neq 00
  \end{cases}
  \\
  \vcpcl_\beta&(T)
  =
  \left.
    -\nabla_{\vpcl_\beta} \Jobj
  \right|_T
  \nonumber\\
  &=
  \begin{cases}
    \left.
      -\Bigl(3\vpcl_{00}-\vpcl_{01}-\vpcl_{10}-\vpcl_{11}\Bigr)
    \right|_T,
    &
      \beta=00
      \\
      \left.
        -\Bigl(\vpcl_\beta-\vpcl_{00}\Bigr)
      \right|_T,
      &
      \beta\neq 00
    \end{cases}
    \\
    \cS_\beta&(T)
    =
    \left.
      -\nabla_{S_\beta} \Jobj
    \right|_T
    \nonumber\\
    &=
    \begin{cases}
      \left.
        -\Bigl(3S_{00}-S_{01}-S_{10}-S_{11}\Bigr)
      \right|_T,
      &
      \beta=00
      \\
      \left.
        -\Bigl(S_\beta-S_{00}\Bigr)
      \right|_T,
      &
      \beta\neq 00
    \end{cases}
\end{align}
where $(-1)^\beta$ is $+1$ for $\beta=00$ and $11$ and $-1$ for
$\beta=01$ and $10$. These equations express the values of the
auxillary varibles at time $t=T$ in terms of the physical varibles
also at $t=T$ and give the input to the backwards propagation of the
auxillary variables, cf.\ Ref.~\cite{krotov95:_global_methods_control}.


\end{document}